\documentclass[utf8]{frontiersSCNS} 

\usepackage{url,hyperref,lineno,microtype,subcaption}
\usepackage[onehalfspacing]{setspace}

\def\keyFont{\fontsize{8}{11}\helveticabold }
\def\firstAuthorLast{Sample {et~al.}} 
\def\Authors{T. M. Garton\,$^{1,*}$, C. M. Jackman\,$^{2,1}$, A. W. Smith\,$^{3}$, K. L. Yeakel\,$^{4}$, S. A. Maloney\,$^{2}$ and J. Vandegriff\,$^{4}$}
\def\Address{$^{1}$Space Environment Physics Group, Department of Physics and Astronomy, University of Southampton, Southampton, England \\
$^{2}$ School of Cosmic Physics, Dublin Institute for Advanced Studies, Dublin, Ireland \\
$^{3}$ Mullard Space Science Laboratory, Department of Space and Climate Physics, University College London, London, England \\
$^{4}$ Johns Hopkins University Applied Physics Laboratory, Laurel, Maryland, United States of America
}
\def\corrAuthor{Tadhg Mark Garton}

\def\corrEmail{t.m.garton@soton.ac.uk}

\begin{document}
\onecolumn
\firstpage{1}

\title[Machine Learning Reconnection Classification]{Machine Learning Applications to Kronian Magnetospheric Reconnection Classification}

\author[\firstAuthorLast ]{\Authors} 
\address{} 
\correspondance{} 

\extraAuth{}

\maketitle

\begin{abstract}

The products of magnetic reconnection in Saturn's magnetotail are identified in magnetometer observations primarily through characteristic deviations in the north-south component of the magnetic field. These magnetic deflections are caused by travelling plasma structures created during reconnection rapidly passing over the observing spacecraft. Identification of these signatures have long been performed by eye, and more recently through semi-automated methods, however these methods are often limited through a required human verification step. Here, we present a fully automated, supervised learning, feed forward neural network model to identify evidence of reconnection in the Kronian magnetosphere with the three magnetic field components observed by the Cassini spacecraft in Kronocentric radial-theta-phi (KRTP) coordinates as input. This model is constructed from a catalogue of reconnection events which covers three years of observations with a total of 2093 classified events, categorized into plasmoids, travelling compression regions and dipolarizations. This neural network model is capable of rapidly identifying reconnection events in large time-span Cassini datasets, tested against the full year 2010 with a high level of accuracy (87\%), true skill score (0.76), and Heidke skill score (0.73). From this model, a full cataloguing and examination of magnetic reconnection events in the Kronian magnetosphere across Cassini's near Saturn lifetime is now possible.

\tiny
 \keyFont{ \section{Keywords:} Machine Learning, Magnetic Reconnection, Planetary Magnetospheres, Magnetotail, Plasmoid} 
\end{abstract}

\section{Introduction}

Magnetic reconnection is the primary process whereby magnetic fields under strain can reconfigure and energy within their structure can transfer. On the dayside, incoming plasma and magnetic fields can reconnect, opening previously closed planetary magnetic field lines. At planets like Earth, day-side (between 6 and 18 local time) reconnection is considered to play a primary role in energy and mass transportation between a planet's magnetic field and the interplanetary magnetic field \citep{Milan07}. Similarly, on the night-side (0 to 6 and 18 to 24 local time), open planetary magnetic field lines become distended in an extended planetary magnetotail, within which field lines may reconnect to again form closed field lines \citep{Dungey61}, \citep{Dungey65}. This cyclic transition between open and closed field configurations allows the transfer of mass, both in and out, of the planetary magnetosphere system. Alternatively, reconnection can further occur for rapidly rotating planets which involves no change in overall magnetic flux. For example, at Jupiter and Saturn fast rotation rates and significant internal mass sources result in the operation of the Vasyliunas cycle.  In this cycle mass is lost down the magnetotail through the reconnection of centrifugally stretched, mass loaded field lines \citep{Vasyliunas83}.

On a global scale reconnection can facilitate an energy balance, dynamic equilibrium between the planetary field and the interplanetary field, and serve as a way to balance the mass budget for magnetospheres where there is a significant amount of internal plasma loading, e.g. from volcanic moons. However, on a small scale it produces local fluctuations of energy and unstable closed magnetic field systems of plasma. These small scale products can be identified by \textit{in-situ} spacecraft through measurements of magnetic field topology and changes in plasma flow. For this study, focus will be on reconnection signatures at Saturn, as identified from the Cassini magnetometer and now classified through machine learning. Reconnection for Saturn has long focused on the planetary magnetotail whereby two types of reconnection signatures are typically reported: dipolarizations and plasmoid ejections. Dipolarizations occur on the planetside of the reconnection site where previously stretched magnetic field lines relax, under a reconnection event, to a more dipole-like magnetic field \citep{Bunce05}, \citep{Russell08}, \citep{Jackman13}, \citep{Jackman15}, \citep{Yao17}, \citep{Smith18a}, \citep{Smith18b}. On the tailside of the reconnection site, closed magnetic field systems encompassing a trapped bubble of plasma known as plasmoids are created during reconnection, which are rapidly ejected down-tail. These events were first identified in Earth's magnetosphere \citep{Hones77}, but have since been identified in Saturn's magnetosphere \citep{Jackman07}, \citep{Hill08}. Observations of these reconnection related structures can be further identified indirectly in magnetic field measurements in the adjacent magnetotail lobes through compressions in the magnetic field. These features are known as travelling compression regions (TCRs; \citet{Slavin84}) due to their close following of plasmoid and dipolarization features. Notably, this indirect method of identification gives no insight to the internal structure of the plasmoid but do at least indicate reconnection occurring and hence can be used to estimate reconnection rates.

\begin{figure*}[t]
\centering
\includegraphics[width=0.99\textwidth]{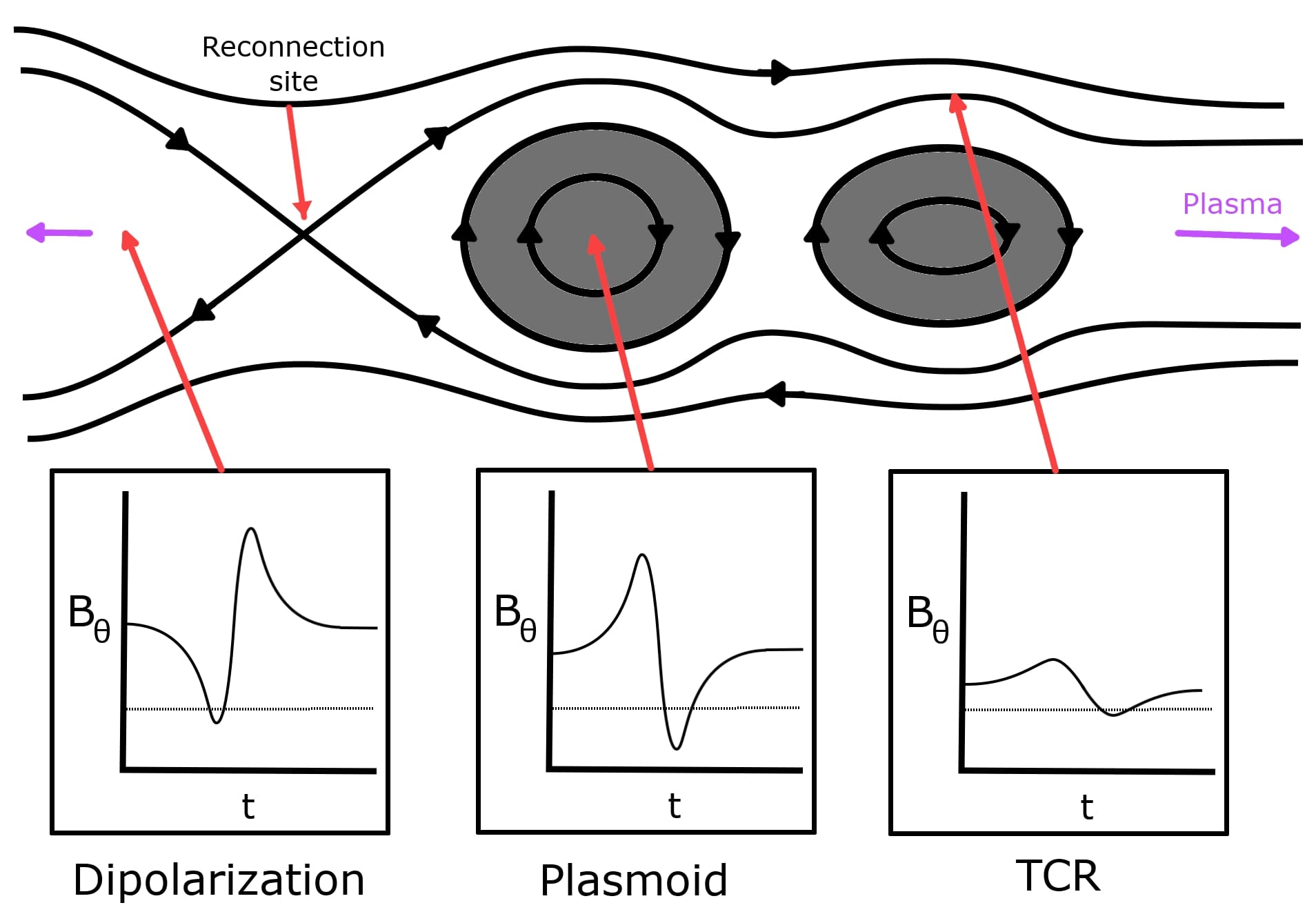}
\caption{Model north-south magnetic field ($B_{\theta}$) measurements for a spacecraft as it passes through a dipolarization, plasmoid and TCR associated with a magnetic reconnection event. Notably, a significant deflection occurs as the spacecraft travels through the center of this region, with directionality of the field even possibly being reversed (going from positive to negative).}\label{Plas_model}
\end{figure*}

Typically signatures of reconnection may be identified by a rapid deflection in the north-south magnetic field component, as observed in Figure~\ref{Plas_model}. At Saturn, plasmoids moving are expected to exhibit a south to north deflection and vice versa (north to south deflection) for planetward-moving dipolarizations. For plasmoids in particular, it is important to recognize the true velocity of the signature may have some azimuthal/corotational component following release \citep{Mcandrews07}, \citep{Thomsen13}, \citep{Neupane19}, \citep{Kane20}. The nature and magnitude of magnetic field deflection depends not only on the intensity of incoming reconnection event, but also on the orientation and direction of travel of the observing spacecraft through this region \citep{Cowley15}. Spacecraft travelling through the center of reconnection signatures observe stronger deviations from the background north-south component of magnetic field, and vice versa. Without \textit{a priori} knowledge of reconnection, these signatures are the principal identifiable feature in magnetic field data, and any deviation in north-south field component present a potential indication of magnetic reconnection. Notably, this is not a definitive method of classification as random turbulent motion in the magnetosphere or waves in the plasma sheet can reproduce similar signatures in the magnetic field observations \citep{Nakagawa89}, \citep{Jackman09}, \citep{Martin17}.

Only recently has there been sufficient data to catalogue and identify large numbers of reconnection events in Saturn's magnetosphere. During 2006 the Cassini spacecraft executed a series of tail orbits to a maximum downtail distance of 68~$R{_S}$ (1~$R{_S}$ = 60268~km) and reconnection signatures from these data were catalogued by \citet{Jackman07}, \citet{Hill08}. These catalogues were built upon in \citet{Jackman11}, where 34 additional plasmoid signatures were identified in the 2006 orbit, and again expanded in \citet{Jackman14} which reported a total of 99 events, 86 of which are identified moving tailward. Estimations of mass loss from large-scale events in this catalogue could not balance the mass gain in the system from Enceladus and other sources \citep{Bagenal11}. Multiple theories have been submitted to account for this imbalance including unobserved mass loss in the magnetospheric flanks \citep{Burkholder17}, \citep{Ma17}, through small scale processes \citep{Bagenal11}, simply that the definition of reconnection event duration under-accounted for the mass in a plasma structure \citep{Cowley15}, or unaccounted for reconnection on the day-side may balance the mass transfer budget \citep{Guo18}. Most recently, \citet{Smith15} attempted to more fully quantify the mass imbalance through the creation of a more comprehensive model and catalogue of tail reconnection events. This model was applied to the equatorial dawn flank orbits and midnight tail orbits of 2006, the dusk flank orbits of 2009, and similar low latitude dusk orbits throughout 2010. Across this observing window 2093 individual events were identified and validated forming a substantial catalogue of reconnection events for Saturn's magnetosphere. However, their semi-automated technique required the selection of observationally defined limits and thresholds.

Here, we apply established methods of machine learning (ML) to planetary magnetospheric reconnection classification to expand these previous surveys to spatially cover the entire Kronian magnetosphere and temporally cover all of Cassini's near Saturn lifetime. ML is an application of artificial intelligence that allows computers the ability to learn from large datasets and experience without being explicitly programmed. This method aides in the prevention of biases and limitations that would otherwise be imposed by a human created model, such as event size and spatial constraints. Furthermore, these models perform well at identifying underlying structures that humans otherwise would not, or could not, that are essential for classification and can be extrapolated to identify features in previously unobserved datasets and have already been implemented across the field of astrophysics to solve a variety of problems \citep{Ruhunusiri18, Ruhunusiri18b, Waldmann19}.

\section{Dataset and Observation}

The datasets used in this study are magnetic field component measurements as observed by the Cassini magnetometer (MAG; \citet{Dougherty04}) instrument. Cassini was launched onboard a Titan IV rocket in 1997 and following Saturn Orbit Insertion (SOI) in July 2004, it orbited the planet until 2017. During its lifetime it observed a variety of environments within the Kronian magnetosphere which can be used to gain a greater understanding of Saturn's magnetic processes. For this research, Kronocentric radial, theta, phi (KRTP) coordinates are used as this coordinate system has been shown to be useful in distinguishing reconnection related events from turbulent motion in the hinged current sheet \citep{Jackman09}. In this spherical coordinate system the radial component ($B_r$) is positive outward from Saturn, the meridional component ($B_{\theta}$) is positive southward (at the equator), and the azimuthal component ($B_{\phi}$) is positive in the direction of corotation (prograde). Furthermore, one minute cadence observations are analyzed as it has been shown that reconnection events last an average duration of $\sim$10-20 minutes and can be accurately identified at this cadence \citep{Jackman14}, \citep{Smith15}.

\begin{figure*}[t]
\centering
\includegraphics[width=0.99\textwidth]{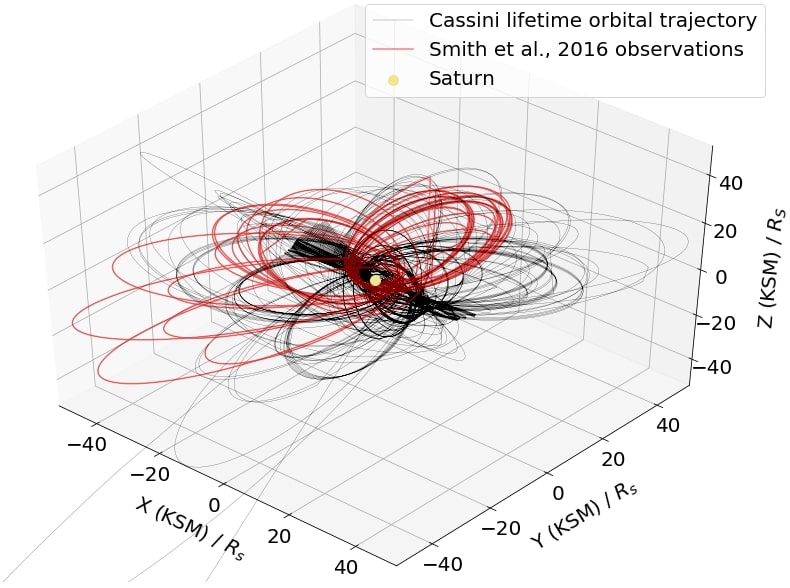}
\caption{Lifetime trajectory (black) of Cassini around Saturn (yellow). The Cassini trajectories during the observing window employed in the creation of the \citet{Smith15} catalogue of magnetic reconnection are highlighted in red for comparison. For the creation of a machine learning training set, observations of events are taken from the Smith catalogue and null events are randomly taken from a variety of local times and radial distances during the Smith catalogue observing window.}\label{Cass_traj}
\end{figure*}

Figure~\ref{Cass_traj} illustrates the near-Saturn lifetime trajectory of Cassini in Kronocentric solar magnetospheric (KSM) coordinates. This Cartesian coordinate system is oriented such that the x axis points toward the Sun, the x‐z plane contains the planetary dipole axis, and the y component completes the right‐handed set. The trajectories of Cassini during the \citeauthor{Smith15} observing window is highlighted in red for comparison. The full 13 years dataset shows the various magnetic environments about Saturn that the Cassini satellite has explored. Similarly, the trajectories during the highlighted observations cover much of these varied environments, however are focused primarily on longer observation times of Saturn's magnetotail within the equatorial plane. Furthermore, this observing window covered times when Saturn's night-side current sheet was hinged upward (southern hemisphere summer), was parallel to the equatorial plane (e.g. equinox; \citet{Khurana09}), or even hinged downward (northern hemisphere summer) later in the mission \citep{Arridge11}. By allowing for identification across the entire Cassini lifetime, more accurate statistical investigations can be performed on reconnection occurrence across the entire morphology of Saturn's magnetosphere.

\begin{figure*}[t]
\centering
\includegraphics[width=0.69\textwidth]{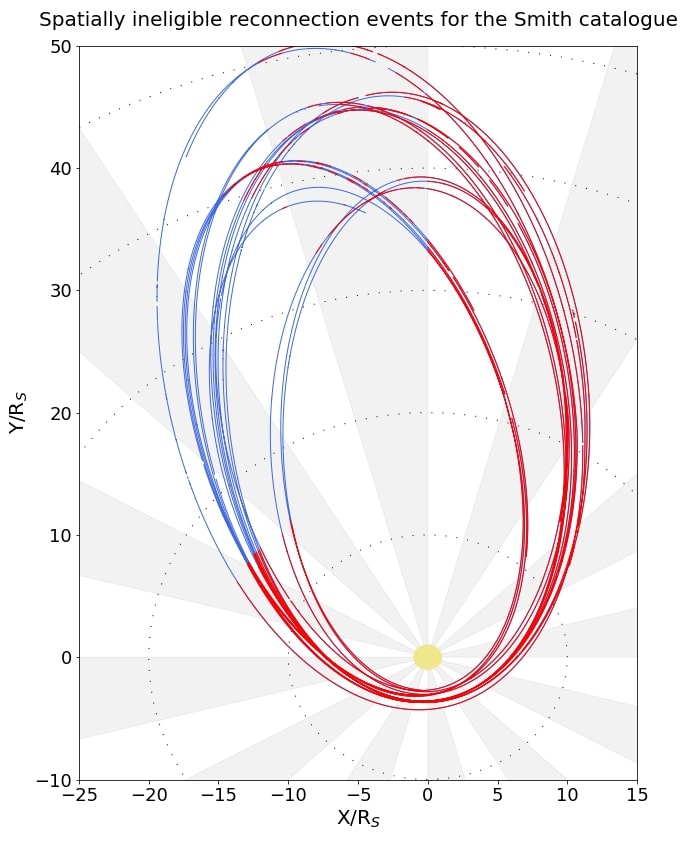}
\caption{2010 trajectory of Cassini about Saturn (yellow) separated by colour into regions where the S16 could identify reconnection events (blue) and the trajectories that were spatially ineligible for identification (red). Notably, at large distances ($>$35~$R{_S}$) eligibility appears to be very patchy, this is due to the changing position of the magnetopause boundary under the varying balance between solar wind dynamic pressure and internal plasma pressure.}\label{inelig_spatial}
\end{figure*}

For the construction of a supervised ML model, a previous, labeled database is required for the model to learn the parametric identifiers of the magnetic reconnection class, and to test against to validate the model's accuracy. The \citet{Smith15} catalogue (hereafter S16) of reconnection is selected as this classified dataset due to its large number of samples, variety of orbital trajectories sampled, and its final human based verification step. However, to utilize this catalogue, the limitations of its selection criteria must be understood. This catalogue was constructed from a semi-automated model with many hard-coded limitations. Excluding the aforementioned temporal limitations of observation window selection, this model further includes spatial and magnetic parametric limitations. Spatially, this model is defined within a 'viewing region' where events are strictly only identified within the night-side, at distances greater than 15~R$_{S}$ from Saturn, and strictly within the magnetosphere. Figure~\ref{inelig_spatial} demonstrates the spatial constraints on the S16. This figure illustrates the entire 2010 trajectory of he Cassini instrument seperated into spatial constraints where the S16 could identify reconnection events (blue) and those where identifications are spatially ineligible (red). This catalogue has similar magnetic parametric limitations. Primarily events are identified from the background through a quadratic fit to $B_{\theta}$ polarity crossings with a least squared goodness of fit value of $r^2 \ge 0.9$. Identified candidates are then verified through

\begin{equation}
\frac{|\Delta B_{\theta}|}{B_{\theta}^{RMS}} \ge 1.5
\label{criteria_1}
\end{equation}

\noindent where $|\Delta B_{\theta}|$ is the magnitude of deflection during the event and the root-mean-square (RMS) of $B_{\theta}$ is calculated for a period extending 30 minutes both sides on the candidate. A secondary validation step follows this such that:

\begin{equation}
|\Delta B_{\theta}| \ge 0.25~nT
\label{criteria_2}
\end{equation}

\noindent where symbols have their previous meaning. These validation steps are imposed as it is difficult for humans to verify candidates that fall below these parametric limitations due to a signal to noise ratio problem. Through these identification and validation methods, the \citeauthor{Smith15} model identifies 2094 (1083 planetward and 1011 tailward) reconnection signatures within their observation window.

These events identify the temporal windows which act as a labeled dataset for a supervised training ML method. However, training of a ML model requires a collection of input parameters, from which the ML model learns the association of parameters to events. For this research, exclusively magnetic observations in the three spatial components of the KRTP coordinate systems are used for identification. This selection is made due to the coverage of Cassini's lifetime that the MAG instrument remained operational. While signatures of planetary reconnection exist in other property observations such as plasma density, MAG data is used as a predominant identifier for human based identifications. Furthermore, the Cassini plasma spectrometer (CAPS; \citet{Young04}) did not remain operational across the entirety of Cassini's near Saturn lifetime, being permanently inactive post-2012, nor did it provide a full 3D picture of the plasma environment, and so may miss any reconnection related jets due to pointing in the ‘wrong’ direction. A model for identifying magnetic reconnection signatures using only magnetic field component data would also ease possible transitions, and transfer learning of a ML model to use with new satellites and for different planetary magnetic fields. Hence, plasma property observations for these reconnection events are not used in this research, however, plasma observations could and should be used in any future implementation where the plasma measurements are comprehensive in both time and 3D viewing. Finally, it is envisioned that the construction of a catalogue using this method across the entire Cassini dataset will enable the examination of numerous case studies of reconnection using multiple instruments.

\begin{figure*}[t]
\centering
\begin{minipage}[c]{0.5\textwidth}
 \includegraphics[width=1\textwidth]{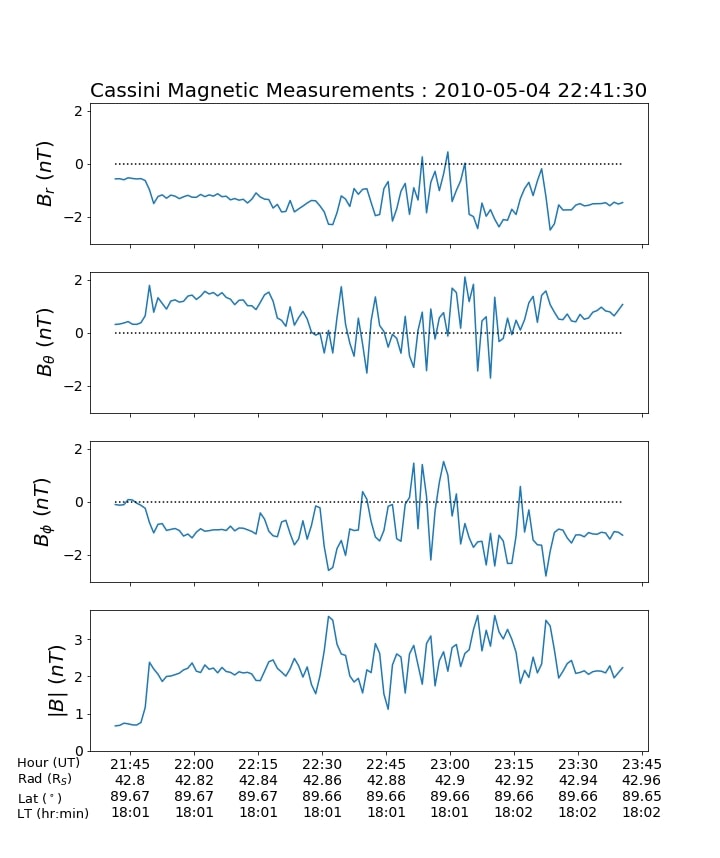}
\end{minipage}%
\hfill
\begin{minipage}[c]{0.49\textwidth}
 \includegraphics[width=1\textwidth]{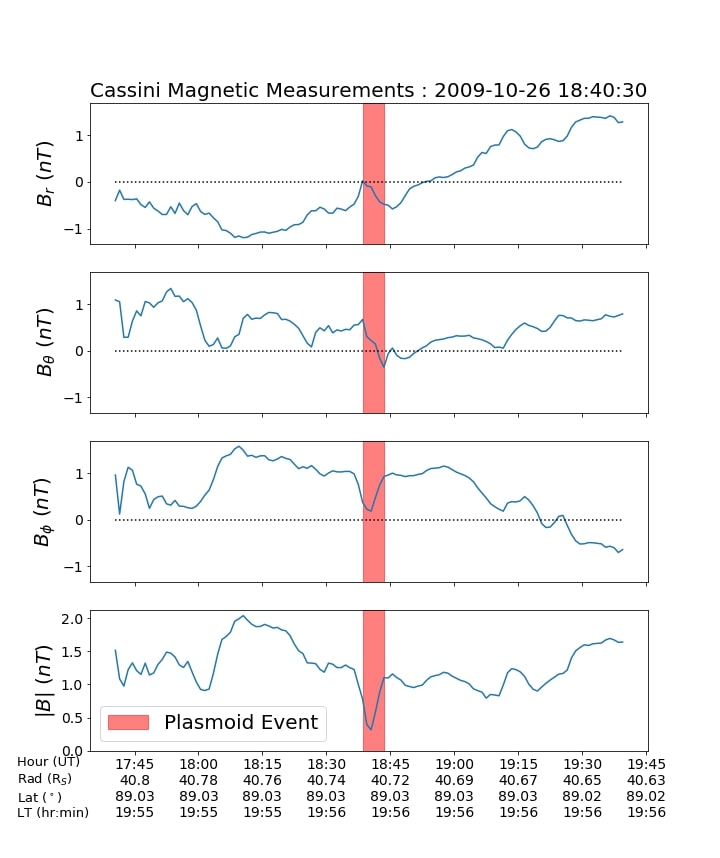}
\end{minipage}%
\caption{Examples of magnetometer data for a non event (left) and event (right) used to train a machine learning algorithm. Titles of these plots denote the time at at center of these observing windows in a YYYY-MM-DD format.}\label{train_comp}
\end{figure*}

Figure~\ref{train_comp} illustrates example magnetic time series across the three KRTP spatial components as well as the total magnetic field, $|B|$, used during training as a null classification (left) and an event classification (right). The X-axis of these plots denotes the time of observation and the spacecraft ephemeris data for Cassini at that time. The time constraint of ML training is highly dependant on the size of input parameters, hence, only the three elementary components of magnetic field measurements from Cassini are used as inputs for ML training in this study.

\begin{equation}
|B| = \sqrt{B_{r}^2 + B_{\theta}^2 + B_{\phi}^2}
\label{B_total}
\end{equation}

\section{Machine Learning Architecture}

\subsection{Class balancing and Data Augmentation}

The greatest risk for poorly constructed ML identification of relatively rare features is the possibility of a class imbalance \citep{Buda17}. For this case, magnetic reconnection events are only identified occupying $\sim<$1-10$\%$ of the total observing time dependant on the identification method, hence, ML training with this ratio will exhibit bias towards the majority class \citep{Gua08}, \citep{Johnson19}. Hence, an unbalanced ratio of non-events to events will cause the ML algorithm, in its interest of maximizing its accuracy, to simply classify all inputs as nulls to obtain an accuracy of $\sim$90$\%$ without truly learning underlying identifying signatures. To alleviate this issue, a randomized under-sampling of non-reconnection events is used to balance with the $\sim$2000 events in the S16. This renders $\sim$4000 total observations to construct training, test and validation sets, which is a low number of samples to perform ML methods to and expect the overarching reconnection features to be accurately identified, rather than the ML model simply memorizing the training set. 

The issue of a small sample size can be solved through data augmentation, such as data synthesis, or the transformation of already existing data \citep{Mikolajczyk18}, \citep{Fawaz18}. Data synthesis is simply the creation of data through the combination of a model with some overlying noise in an attempt to create real-like datasets, however this method can be inaccurate if predictive models are inaccurate, or missing some underlying understanding. Data transformation takes already existing data and applies some kind of transformation, such as adding noise or filters over the existing measurements or translating the data either spatially or temporally. Since the signatures of magnetic reconnection occur across a number of minutes, averaging $\sim$8 minutes \citep{Smith15}, it is possible to increase our number of samples by considering every minute of an event as a unique positive identification. Hence, a single event lasting 5 minutes would be considered as 5 consecutive positive labelled identifications every minute between the start and end time of an event. This method increases the total available observations to $\sim$32000 (16000 positive labels and 16000 randomly selected negative labels). This increased number of samples allows for more complex ML architectures and a more robust final model. In this instance, nulls are selected randomly from the S16 observing window with the same spatial limitations of the S16, e.g. at distances greater than 15~$R_S$ from Saturn, etc. Finally, since these events occur and are identified across multiple minutes of magnetic data, due to their temporal structure, for the ML model to identify these events, it must have a time window of magnetic measurements as input. 15 minutes both before and after the central label in the three KRTP spatial magnetic field components ($B_r, B_{\theta}, and B_{\phi}$) are used as this window is wide enough to cover the longer duration events in the S16 catalogue, but short enough to identify label changes occurring between event clusters. This renders a total of 90 magnetic property inputs for each of the 32000 labels for any given ML model.

\subsection{Machine Learning Types}

A variety of ML models exist, ranging in complexity to allow for identification of more elaborate and subtle features within datasets. This research focuses on identification of features within three singular dimension magnetic field time series, hence, only relatively simple supervised learning ML methods will be investigated, namely: support vector classifier with a linear (LSVC) and non-linear kernel (NLSVC), random forest classifier (RFC), and a simple artificial feed forward neural network (ANN). All of these models are available in the sklearn python packages \citep{Buitinck13} and the TensorFlow libraries \citep{TensorFlow15}. A LSVC creates a multi-dimensional hyperspace of observed parameters. The labeled data are then input into this hyperspace and a linear hyperplane is created as a decision boundary to optimally separate data of opposing labels with the widest possible margins. This hyperplane seperator is then stored and used to predict the labels of new datasets. A NLSVC behaves similarly to its linear variant, by creating some hyperplane as a decision boundary, however, the kernel function utilized by a NLSVC can non-linearly transform the feature space such that the classes become separable. RFC similarly creates a multi-dimensional hyper space, but instead of separating data by a continuous hyperplane, a vast array of boolean decision tree networks, of variable depth, are created to segment a training dataset non-linearly. New data sets are then input into this array of decision trees and a classification is judged by majority vote outcome. The final type, ANNs, rely on the creation of input (parameters) and output (labels) neural nodes, interconnected by a collection of initially random weights and biases. This method of ML is optimized through tuning of various hyperparameters such as: the non-linear activation function on each of the nodes, the number of nodes within each layer, the loss and optimization functions, and the number of hidden layers within the architecture. These hidden layers of neural nodes between the input and output nodes have no true observable parameter, however they enable more complex feature identification by the ANN. To judge which of these models is optimal for identification of reconnection signatures, each must be trained and the model that exhibits the highest accuracy can be selected for further fine tuning. It is important to note that model accuracy is not typically the greatest indicator of a model's performance, and many other metrics will be discussed later, however this metric is significant enough to indicate a single ML model that can be best improved, and hence will be further investigated in this research. Table~\ref{ML_acc} indicates the accuracy for these four ML models to identify the signatures of magnetic reconnection using only the three KRTP magnetic field components observed by Cassini for times within the spatial and temporal limitations of the S16. Overfitting of these models was prevented by standard methods of train/test/validation splitting, principle component analysis and algorithm complexity limitations. The train/test/validation split had a weighted random assignment across all years in the S16 catalogue with no temporal disjoint. This means the training set was composed of events from 2006, 2009, and 2010 allowing it to learn the structure of reconnection from varied spacecraft orbits and trajectories. However, set assignment was performed on a reconnection event basis, meaning all minutes of observations associated with an individual reconnection event are assigned to a single set. Most notably, ANNs exhibit the highest accuracy rating, likely due to their allowed higher complexity when compared to the other methods mentioned. Hence, ANNs are further utilized for this research.

\begin{table}
\caption{Comparison of validation set accuracy for ML event classification using a linear support vector classifier (LSVC), a non-linear support vector classifier (NLSVC), a random forrest classifier (RFC), and artificial neural network (ANN). This accuracy value is calculated as the ratio of correctly identified samples to total samples.}\label{ML_acc}
\centering
\begin{tabular}{l c}
\hline
ML type  & Accuracy  \\
\hline
LSVC  & 0.73   \\
NLSVC  & 0.75   \\
RFC  & 0.87   \\
ANN  & 0.90   \\
\hline
\end{tabular}
\end{table}

\subsection{Artificial Neural Networks}

\begin{figure*}[t]
\centering
\includegraphics[width=0.99\textwidth]{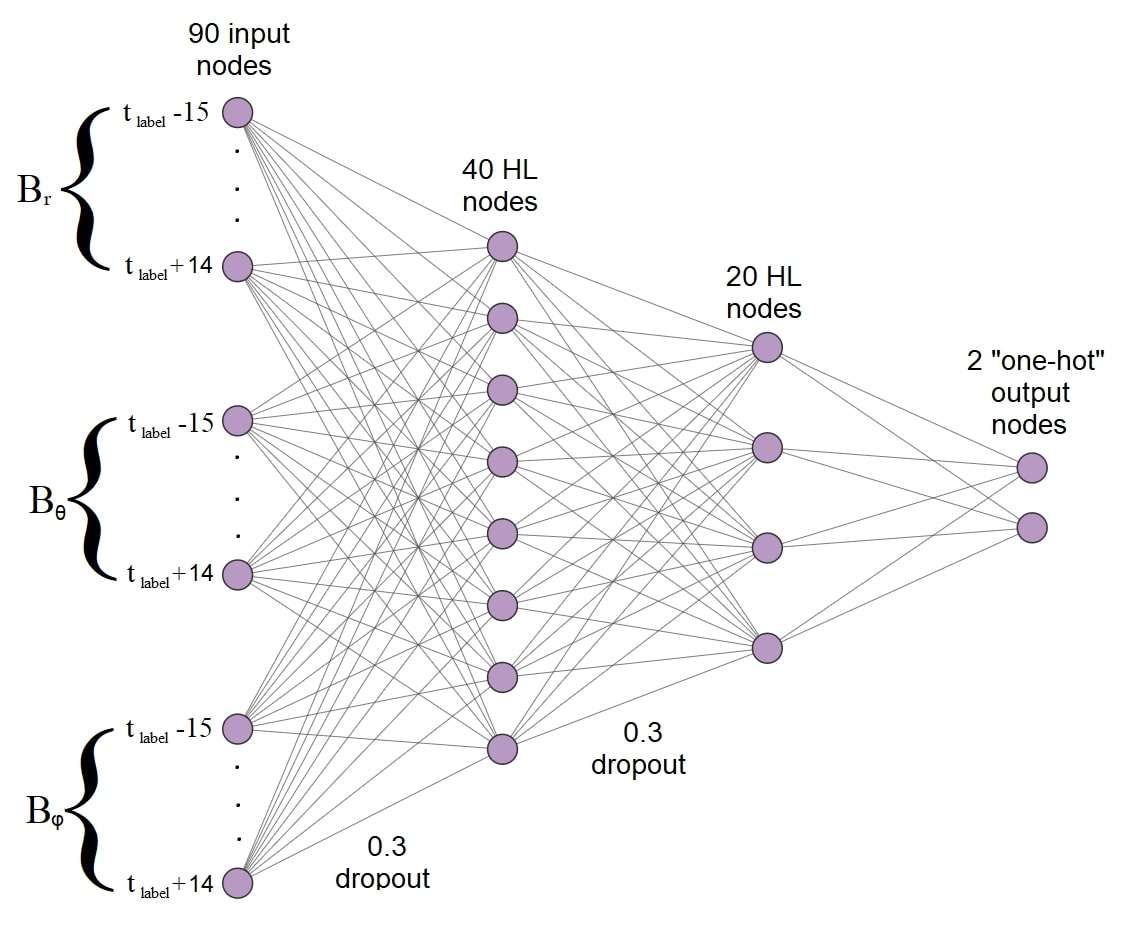}
\caption{NN architecture used to train to identify reconnection signatures in Cassini magnetometer data. This structure shows 90 input nodes composed of three 30 minute time windows centered on the label time ($t_{label}$), in the three KRTP magnetic field components ($B_r$, $B_{\theta}$, and $B_{\phi}$). These nodes are then fed into a 40 node hidden layer (HL) with a 0.3 dropout,  which feeds into a 20 node HL with a 0.3 dropout. This final HL is then categorized using a 2 node, one-hot classification system. During training, every epoch, the weights and biases interconnecting each layer are varied to under a gradient descent to optimize the accuracy of classifications.}\label{NN_architecture}
\end{figure*}

Figure~\ref{NN_architecture} demonstrates the architecture of a simple ANN created and trained during this research to identify signatures of magnetic reconnection. In this architecture, input properties are directed into the architecture in the input layer. Operations are performed on these parameters between each interconnected layer, with the goal being to accurately recreate the desired outputs in the output layer. ANNs are generally optimized and fine tuned through a process of trial and error, however some simple rules for their creation exist to prevent overfitting of training data. Generally, the number of free parameters must not exceed the number of samples used for training, i.e.

\begin{equation}
N_{S} > N_{FP} = \sum_{i=1}^{i}\Big((N_{i-1}\times N_i) + N_{i}\Big)
\label{NN_FP}
\end{equation}

\noindent where $N_S$ is the number of training samples, $N_{FP}$ is the number of free parameters, and $N_{i}$ describes the number of nodes in the ith layer. No strict consensus exists to decide the number of nodes in ANN hidden layers, however it is generally accepted for the number of nodes in a hidden layer to be approximately half way between the number of nodes in the previous and next layers. Through trial and error, it was found that a two hidden layer ANN architecture was most efficient at identifying magnetic reconnection in the training set, however \citet{Huang03} proved an upper limit to the total available hidden nodes available in this system to be

\begin{equation}
N_{H} \leq \frac{2}{\alpha}\sqrt{(N_O+2)N_S}
\label{num_hl_nodes}
\end{equation}

\noindent where $N_S$ has its previous meaning, $N_H$ represents the total available hidden nodes, $N_O$ is the number of output nodes, and $\alpha$ is a robustness factor usually between one and ten. From equations~\ref{NN_FP} and \ref{num_hl_nodes}, and the aforementioned 32000 samples, it is possible to train the robust two hidden layer neural network in Figure~\ref{NN_architecture}: 90 input nodes with a dropout of 0.3 connected by a rectified linear units (relu) activation function to 40 first hidden layer nodes, which are in turn connected with a dropout of 0.3 and a relu activation function to 20 second hidden layer nodes, which connects fully with a softmax activation function to two output nodes representing a boolean classification of reconnection occurring. After each training epoch, the model was trained towards improving a binary cross entropy accuracy metric. During training, however, it was observed that a significant number of events were identified outside the magnetosphere, along portions of Cassini's orbit in the magnetosheath and solar wind. This is likely due to the ML algorithm never encountering observations from these magnetic regions during training. Since these regions are unique classifications and differ from null training samples within the magnetosphere, they can be included in training as a unique classification of nulls. This means our number of samples will increase to $\sim$16000 reconnection events, $\sim$16000 magnetosphere nulls, $\sim$16000 magnetosheath nulls, and $\sim$16000 solar wind nulls. Given a train-test-validation split of 60-20-20, $\sim$38400 samples are available for training. 

\begin{figure*}[t]
\centering
\includegraphics[width=0.99\textwidth]{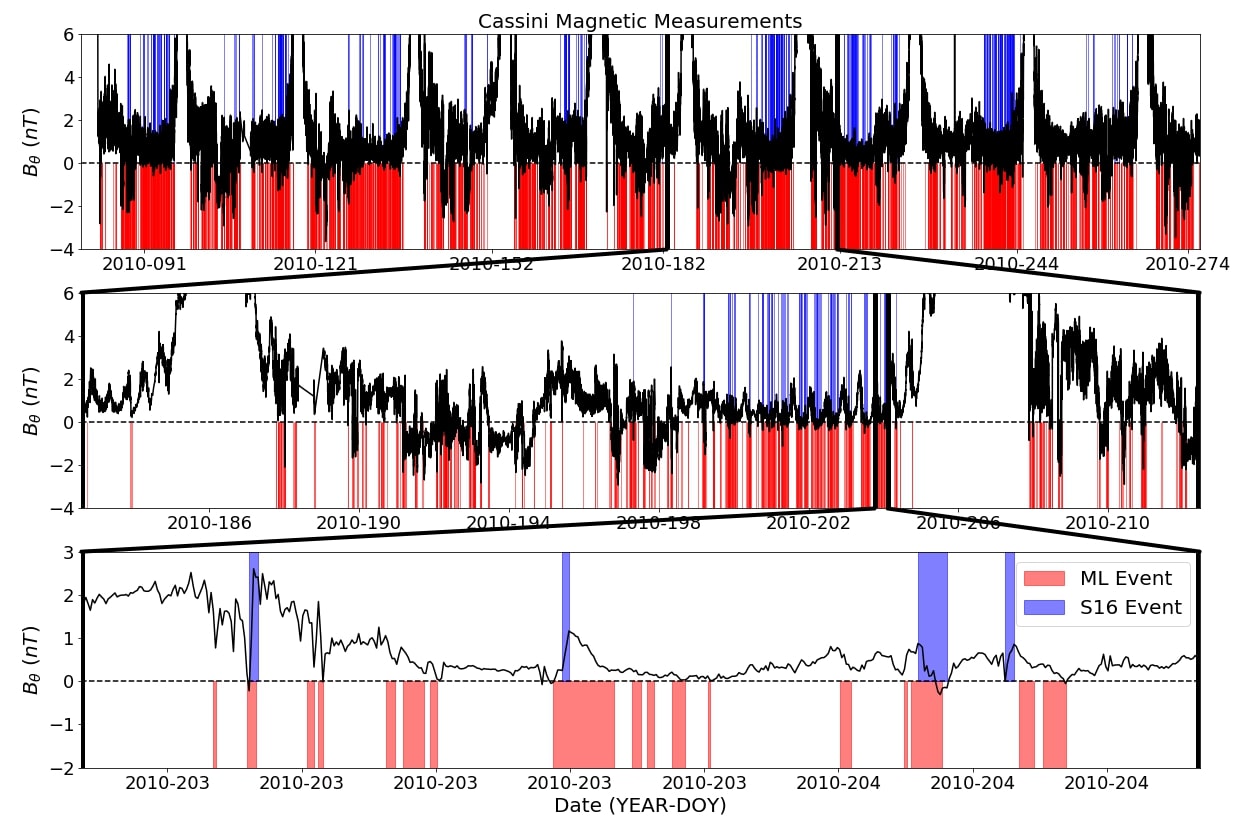}
\caption{Output of reconnection signatures identified by a feed forward neural network (red areas) across half of 2010 compared to identifications from the Smith catalog (blue areas) for the same period. These areas are overplot onto the B$_{\theta}$ component of the magnetic field, where reconnection signatures are easiest identified by eye. Each successive plot examines zoomed in windows to observe finer structure in magnetic field measurements and identifications.}\label{full_year_output}
\end{figure*}

The relative effectiveness of this architecture is displayed in Table~\ref{C_matrix_1} through four confusion matrices. A confusion matrix exists for each of the training, test and validation set, and a fourth confusion matrix illustrates the effectiveness of the ANN to identify reconnection events across the entirety of 2010, replicating how the model will perform on large continuous datasets. The year 2010 was selected for this comparison as it is one of two full years which the S16 covered, along with 2006. 2010 was selected between these two years as the trajectory of Cassini for this year included a wider sampling of varied magnetic environments, hence being the most stringent full year comparison possible. It is important to recognize that this 2010 confusion matrix includes identifications from the training, test, and validation datasets. Across each of these confusion matrices an accuracy of $\sim$90$\%$ is attained and the training, test, and validation sets have high skill metrics: the Heidke skill score (HSS; 0.75; \citet{Heidke26}), the true skill statistic (TSS; 0.76), and the threat score (TS; 0.68). It is important to reinstate, the final step of the S16 catalogue’s final step is a human verification, hence our comparison in the validation confusion matrix shows the effectiveness of the ML model against human verified data. However, in the 2010 confusion matrix, the number of false positives (FP; 32954) significantly outweigh the number of true positives (TP; 5111) leading to a high false alarm ratio. Hence the imbalance in this confusion matrix is represented in its HSS; 0.21, TSS; 0.75, and TS; 0.13. These skill score metrics quantifiably describe the ability of this model to replicate the observable data. The HSS measures the fractional improvement of the forecast over a standard forecast and ranges from $-\infty$ to 1, with 1 being perfectly skillful, a value of 0 representing no skill, and a value of 0.3 being considered of good skill. The TSS, also known as the Peirce's skill score, compares classification to a random selection classifier and ranges from $-$1 to 1, with 1 being considered perfectly skillful, and 0 having no skill. TS measures the fraction of observed and/or classified events that were correctly identified and ranges from 0 to 1, with 0 having no threat detecting capabilities and 1 being a perfect identifier. The imbalance of these classifications is illustrated in Figure~\ref{full_year_output} which compares identification of magnetic reconnection across 2010 by the ANN architecture compared to the S16. In this figure, events are highlighted over underlying $B_{\theta}$ magnetic components as measured by Cassini. Events from the S16 are highlighted in blue, whereas events classified by the ML algorithm are highlighted in red.

\begin{table}
\caption{Confusion matrices for a feed-forward neural network classification of magnetic reconnection within the Kronian magnetosphere}\label{C_matrix_1}
\centering
\begin{tabular}{|p{1.9cm}||p{2.2cm}|p{2.2cm}||p{2.4cm}|p{2.2cm}|}
 \hline
  & \multicolumn{2}{c||}{\textbf{Train}} &  \multicolumn{2}{|c|}{\textbf{Test}}\\
 \hline
  & \cellcolor{blue!25}\textit{Pred. Null} & \cellcolor{blue!25}\textit{Pred. Event} & \cellcolor{blue!25}\textit{Pred. Null} & \cellcolor{blue!25}\textit{Pred. Event}\\
 \hline
  \cellcolor{yellow!25}\textit{Obs. Null} & \cellcolor{green!10}26690 (0.92) & \cellcolor{red!10}{2278 (0.08)} & \cellcolor{green!10}3272 (0.94) & \cellcolor{red!10}226 (0.06)\\
 \hline
 \cellcolor{yellow!25}\textit{Obs. Event} & \cellcolor{red!10}1564 (0.16) & \cellcolor{green!10}8092 (0.84) & \cellcolor{red!10}672 (0.19) & \cellcolor{green!10}2826 (0.81)\\
 \hline
 \hline
  & \multicolumn{2}{|c||}{\textbf{Validation}} & \multicolumn{2}{|c|}{\textbf{2010}}\\
 \hline
  &\cellcolor{blue!25}\textit{Pred. Null} & \cellcolor{blue!25}\textit{Pred. Event} &  \cellcolor{blue!25}\textit{Pred. Null} & \cellcolor{blue!25}\textit{Pred. Event}\\
 \hline
 \cellcolor{yellow!25}\textit{Obs. Null} & \cellcolor{green!10}3020 (0.93) & \cellcolor{red!10}232 (0.07) & \cellcolor{green!10}486400 (0.94) & \cellcolor{red!10}32954 (0.06)\\
 \hline
 \cellcolor{yellow!25}\textit{Obs. Event} & \cellcolor{red!10}614 (0.19) & \cellcolor{green!10}2638  (0.81) & \cellcolor{red!10}1134 (0.18) & \cellcolor{green!10}5111 (0.82)\\
 \hline
\end{tabular}
\end{table}

\section{Discussion}

\begin{figure*}[p]
\centering
\includegraphics[width=0.98\textwidth]{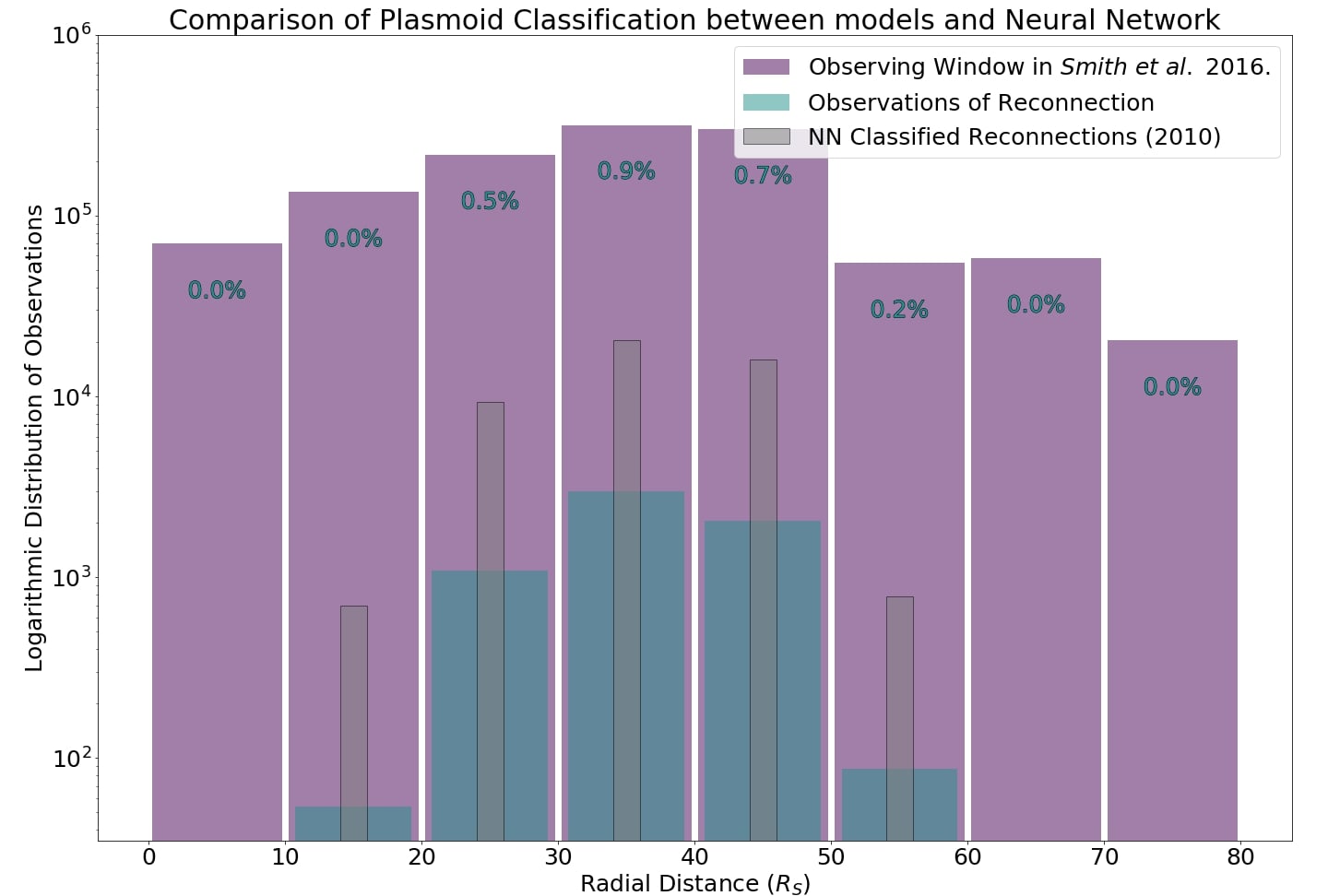}
\begin{minipage}[c]{0.49\textwidth}
 \includegraphics[width=0.99\textwidth]{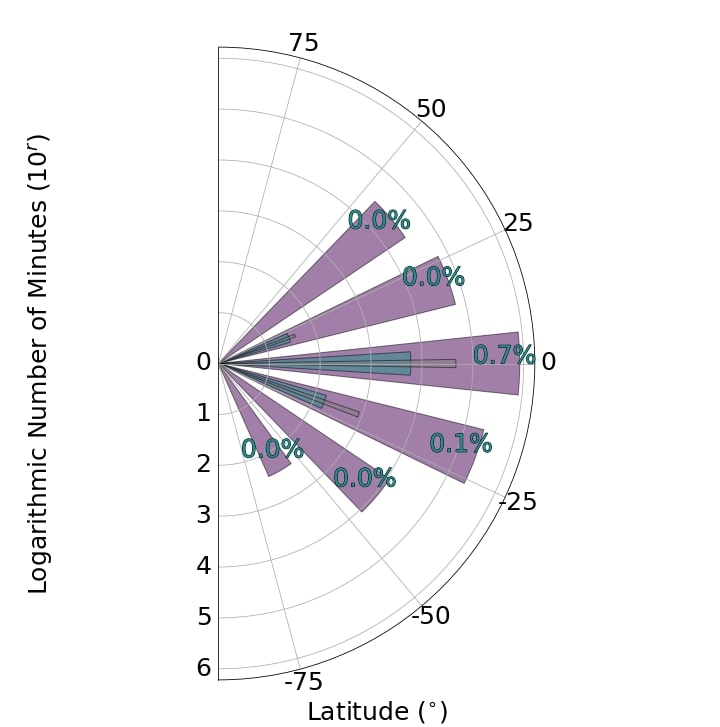}
\end{minipage}%
\hfill
\begin{minipage}[c]{0.49\textwidth}
 \includegraphics[width=0.99\textwidth]{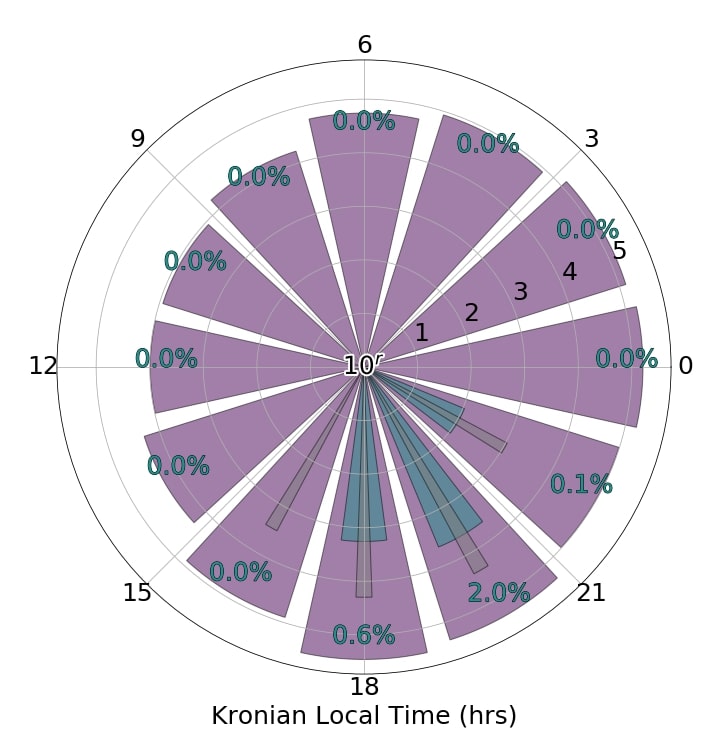}
\end{minipage}%
\caption{Total time of Cassini observations of Saturn's magnetosphere during the 2010 observing window (purple) with radius, latitude and local time respectively. This distribution is compared to the time classified as magnetic reconnection signatures by \citeauthor{Smith15} (blue) and as classified by a neural network method (grey). Percentiles indicate relative time spent near reconnection events as found in the \citeauthor{Smith15} catalogue to the total window.}\label{NN_comp}
\end{figure*}

The results and corresponding skill scores from Table~\ref{C_matrix_1} would imply a significant bias of the neural network to mis-classify null observations, as classified by the S16 catalogue, as events. Investigations into the spatial distribution of events to identify the cause of this large number of mis-classification are illustrated in Figure~\ref{NN_comp}. This figure demonstrates the distribution of total time during the observation window of \citet{Smith15} (purple) across radial distance, latitude and the Kronian local time. Additionally, the time spent observing reconnection related events as stated by the S16 (blue) and the time spent observing reconnection products as classified by the ANN (gray) are displayed for comparison. Blue percentile values illustrate the percentage of total time of a given distribution spent observing reconnection as found by S16. As is illustrated, the ANN observations have a similar spatial distribution of identifications to the S16, simply the ANN recognizes more minutes of reconnection occurring due to more events being identified. In the local time distribution, all events identified by both S16 and the ANN for 2010 are located on the planetary dusk side due to the orbital trajectory of the Cassini spacecraft at this time, being very close to the planet ($<$15$R_S$) at other local times. Most notably, the local time distribution of the ANN identifications shows a non-zero rate of reconnection on the day-side of Saturn, while the \citeauthor{Smith15} model maintained a strict cut-off of dayside events due to its hard coded spatial limitations. Evidence for dayside reconnection has been identified previously \citep{Delamere15}, \citep{Guo18}, hence, inclusion of dayside reconnection identification within this catalogue allows for more future exploratory research to be performed. 

\subsection{Evaluation of ANN Performance and Identifications}

As previously mentioned, the S16 is constructed from numerous hard coded spatial and magnetic limitations within their semi-automatic identification method that significantly limit their identifications. In the ML model, these limitations are not in place, which leads to a substantial number of ML identifications that cannot otherwise be identified by the S16 method, thus leading to our abundance of apparent FPs. Hence, the confusion matrix for 2010 in Table~\ref{C_matrix_1} does not accurately compare the results of the neural network to the S16, and it must be corrected. By examining only the neural network reconnection identifications that could be recognized by the S16 (i.e. events with $\delta B_{\theta}\ \ge\ 0.25\ nT$, and a significant signal to noise ratio: $\delta B_{\theta}/B_{rms}\ \ge\ 1.5$), and comparing events as a whole, by considering sequential positive minute-by-minute classifications as part of the same event, a new confusion matrix is obtained for the entirety of 2010. Table~\ref{C_Matrix_2} demonstrates the corrected confusion matrix for 2010, only comparing events that the S16 could identify. This enables us to more fairly assess the performance of our approach. To calculate the value of true negatives (TN; 1008), the same method could not be used as TN measurements are not considered discrete events, and are not privy to the same parametric limitations that events are. To obtain this value, TNs are considered to be all of the periods when a TP, FP, or FN is not applicable, hence:

\begin{equation}
TN = TP + FP + FN + 1 = 1008
\label{TN_calc}
\end{equation}

\begin{figure}[t]
\centering
\includegraphics[width=0.99\textwidth]{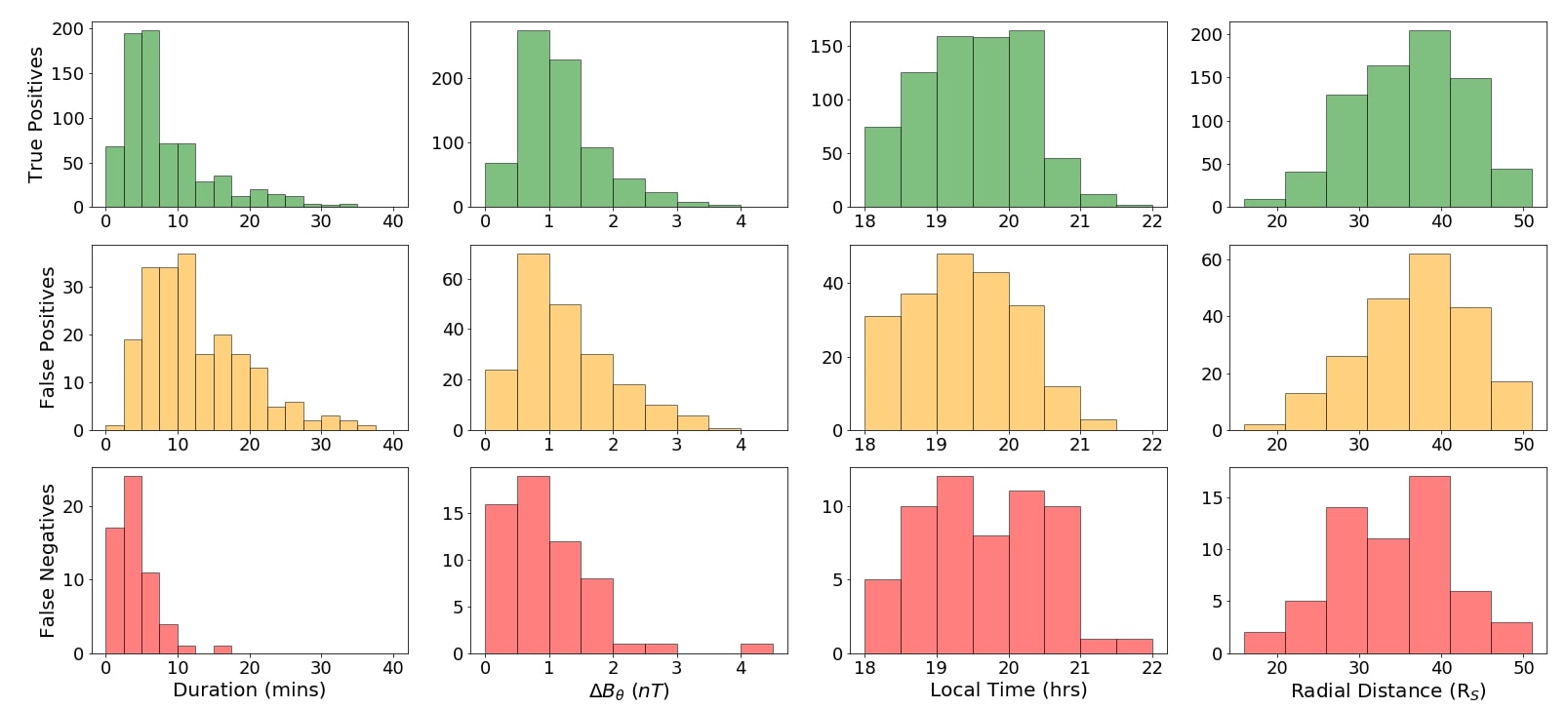}
\caption{Temporal, magnetic and spatial properties of reconnection events that are classified as true positives (green), false positives (orange), and false negatives (red) when comparing the neural network classifications to those of the S16.}\label{Conf_M_props}
\end{figure}

\noindent This corrected confusion matrix for eligible 2010 events has a significant increase in accuracy (87.0$\%$), HSS (0.73), TSS (0.76), and TS (0.74). Figure~\ref{Conf_M_props} displays distributions of temporal (duration), magnetic ($B_{\theta}$ deflections), and spatial (radial distance and local time of event) properties of TP, FP, and FN events from Table~\ref{C_Matrix_2}. No significant discrepancy is evident between these categories spatially or magnetically, however, the differences between the ANN and \citeauthor{Smith15} method is visible in the distributions of event duration. The ANN identifies a higher number of longer duration events, while finding difficulty in identifying short duration events ($<$~10 minutes). However, as evident by the distribution of $\Delta B_{\theta}$ for FNs, these missed events represent smaller deflections, which are least likely to be identified by eye, and most likely to be spurious identifications. The plotted distribution of FPs is very similar to TPs, excluding the longer average durations ($\sim10$ minutes). This discrepancy may be due the quadratic fitting and identification method of the \citeauthor{Smith15} model, coupled with their model not identifying the inclining and declining phases of reconnection which implies a shorter average duration of identifications. Hence, the neural network is considered to accurately identify magnetic reconnection events solely from magnetic field component measurements, not only under the same restrictions as the S16, but also across the total spatial and magnetic domain of Cassini's lifetime.

\begin{table}
\caption{Confusion matrices of neural network classification considering only events that the \citeauthor{Smith15} catalogue could have identified}\label{C_Matrix_2}
\centering
\begin{tabular}{|p{1.9cm}||p{2.2cm}|p{2.2cm}|}
 \hline
  & \multicolumn{2}{c||}{\textbf{2010}}\\
 \hline
  & \cellcolor{blue!25}\textit{Pred. Null} & \cellcolor{blue!25}\textit{Pred. Event}\\
 \hline
  \cellcolor{yellow!25}\textit{Obs. Null} & \cellcolor{green!10}1008 (0.82) & \cellcolor{red!10}208 (0.17)\\
 \hline
 \cellcolor{yellow!25}\textit{Obs. Event} & \cellcolor{red!10}58 (0.07) & \cellcolor{green!10}741 (0.93)\\
 \hline
\end{tabular}
\end{table}

\begin{figure*}[t]
\centering
\includegraphics[width=1\textwidth]{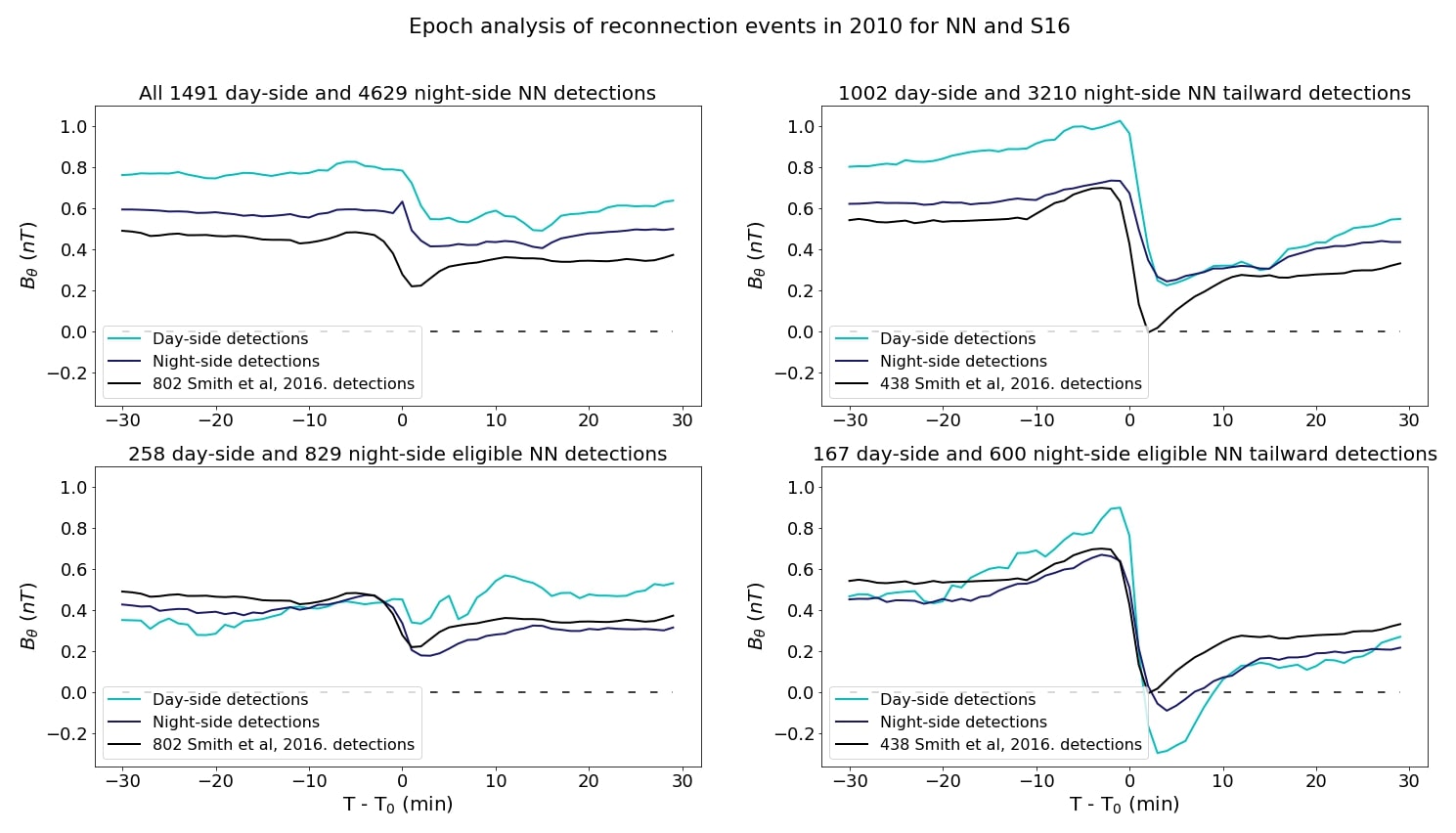}
\caption{Epoch analysis of all 2010 events (top left), tailward events (top right), all 2010 events that meet the S16 criteria (bottom left), and meet the S16 criteria while also being tailward (bottom right) identified by the NN. Identifications are split onto the day-side (light blue) and night-side (dark blue) and are compared to the average of events from the S16 for 2010. (left) and tailward.}\label{epoch}
\end{figure*}

Figure~\ref{epoch} displays an epoch analysis for events classified by this NN for both day-side (light blue) and night side detections (dark blue) compared to the events from the S16 (black). These events are compared across 4 criteria: all events for 2010 (top left), all tailward event for 2010 (top right), all event for 2010 that all within the human built thresholds for S16 (bottom right) and all tailward events that fall within this threshold (bottom right). The term tailward here is defined as a reconnection event occurring with a negative slope in the deflection phase ($B_{\theta}(t_0) > B_{\theta}(t_1)$). The average day-side and average night-side epochs are similar in all panels. The main difference between the two is the higher average $B_{\theta}$ in the day-side events and the larger $\Delta B_{\theta}$ deflections, however this is more likely due to the Cassini spacecraft being closer to the planet on the day-side on average for 2010, and hence within a stronger magnetic field region. The ANN epochs have a similar structure compared to the \citeauthor{Smith15} epoch, however the ANN epochs do not become negative auntil the S16 criteria is applied. This is likely due to the more numerous small scale $B_{\theta}$ deflections ($\Delta B_{\theta} < 0.5$~nT) occurring within a relatively strong magnetic field regions ($B_{\theta} > 1$~nT) for the ANN method than the S16 model, which skews the average. Similarly, events identified by the ANN have higher average $B_{\theta}$ than events identified by the S16, however this is likely due to the ANN not spatially limiting its detections. Interestingly, a secondary deviation is visible in both top panels (no limitations on identifications) at T$\approx$12 minutes after the central deviation. This deviation may imply a propensity for reconnection events to occur in clusters with a $\sim$12 minute delay. However, it is uncertain if this secondary deviation is simply a statistical anomaly in the data, or if this $\sim$12 minute delay is related to the orbital trajectory of Cassini for 2010, particulary since this feature is not visible in the bottom panels (S16 limits in place).

\section{Conclusions}

Here, the operations and effectiveness of ML approaches to magnetic reconnection identification have been discussed. A new ANN model has been constructed to identify reconnection signatures in Saturn's magnetosphere through spherical magnetic field measurements with a HSS$\sim$0.73, and hence is considered an effective identifier. This ML approach identifies deflections in the $B_{\theta}$ field component with no hard-coded limitations that a human-built model may otherwise impose and can identify small scale $B_{\theta}$ deflections that a human, or human made model, would find difficult. This new model has been used across the entire Cassini near-Saturn lifetime to identify $\sim$46000 reconnection events and their associated properties which have been compiled and catalogued. This model and associated reconnection catalogue is available at \citet{Garton20}.

Further study is required on events within this catalogue to identify statistical properties and spatial likelihood of magnetic reconnection in Saturn's magnetosphere to improve predictive modeling. The 13 years catalogue created from this research can be used to identify long-term magnetospheric trends and create a statistical predictive model of reconnection occurence for extreme and rare events. This ANN was constructed using a limited sample of events ($\sim$2000) which may be insufficient to cover the spectrum of reconnection signatures, hence this model can be further improved through the inclusion of additional samples of manually selected reconnection signatures, or through the inclusion of additional particle property observations, should they be available. Furthermore, the training of this ANN involved the inclusion of additional null sets which corresponded to non-reconnection events within the magnetosphere, the magnetosheath and the solar wind. It is possible other such unique magnetic environments exist that could cause spurious identifications where characteristic magnetic field deflections are observed, such as during a Cassini flyby of Titan \citep{Sven10} or Enceledus \citep{Dougherty06}. Hence, inclusion of datasets within these environments as nulls in the training set could improve the overall accuracy and skill of the ANN. Finally, through transfer learning, it is possible to retrain this model to identify similar reconnection signatures in other planetary magnetospheres given fewer training samples of identification. Through this established method it is possible to create a similar operational ML model to identify reconnection signatures at Mercury, or Earth. It is our intention to explore such approaches in future, to realise the full capability of ML for uncovering reconnection signatures for a variety of planetary magnetospheres. Datasets observing various planetary magnetospheres is abundant, e.g. MESSENGER \citep{Solomon01} at Mercury, and Galileo\citep{Young98}/Juno\citep{Bagenal17} at Jupiter, however, exploration of these datasets has only been partially completed by the wider community. This insufficient exploration is partly due to the required time to manually investigate the datasets and the lack of manpower available. ML infrastructure, of the kind discussed in this paper, will enable the processing and full exploration of these large datasets with minimal required human intervention. Furthermore, ML identification methods allow the extrapolation of catalogues and allow for an investigation of more diverse events at different locations, and even make more accurate estimations of the mass budget of magnetospheres. As we rapidly approach a period of data flooding, developing tools to address this issue before it arises is essential for the future of planetary research \citep{Azari20}.

\section*{Conflict of Interest Statement}

The authors declare that the research was conducted in the absence of any commercial or financial relationships that could be construed as a potential conflict of interest.

\section*{Author Contributions}

T. M. G was responsible for development and application of machine learning methods to create a neural network reconnection classifier

\noindent C. M. J. sourced datasets and applied physical insights into the model's creation

\noindent A. W. S. applied physical insights into the model's creation and laid groundwork for its creation

\noindent K. L. Y validated machine learning methods and applied insight into machine learning theory

\noindent S. A. M validated machine learning methods and applied insight into machine learning theory

\noindent J. V validated machine learning methods and applied physical insights into the model's creation 

\section*{Funding}
C. M. J.'s work at Southampton was supported by the STFC Ernest Rutherford Fellowship ST/L004399/1.

\noindent C. M. J.'s work at DIAS was supported by the Science Foundation Ireland Grant 18/FRL/6199.

\noindent T.M.G.'s work is supported by the Science and Technology Facilities Council Opportunities Fund Grant ST/T002255/1.

\noindent A. W. S was supported by STFC Consolidated Grant ST/S000240/1 and NERC grant NE/P017150/1 

\section*{Data Availability Statement}
Calibrated data from the Cassini mission are available from the NASA Planetary Data System at the Jet Propulsion Laboratory [https://pds.jpl.nasa.gov/].

The reconnection catalogue used this study can be found on Zenodo [DOI: 10.5281/zenodo.3978252].

\bibliographystyle{frontiersinSCNS_ENG_HUMS} 
\bibliography{bib}

\begin{thebibliography}{54}
\providecommand{\natexlab}[1]{#1}
\expandafter\ifx\csname urlstyle\endcsname\relax
  \providecommand{\doi}[1]{doi:\discretionary{}{}{}#1}\else
  \providecommand{\doi}{doi:\discretionary{}{}{}\begingroup
  \urlstyle{rm}\Url}\fi
\providecommand{\selectlanguage}[1]{\relax}
\providecommand{\bibAnnoteFile}[1]{%
  \IfFileExists{#1}{\begin{quotation}\noindent\textsc{Key:} #1\\
  \textsc{Annotation:}\ \input{#1}\end{quotation}}{}}
\providecommand{\bibAnnote}[2]{%
  \begin{quotation}\noindent\textsc{Key:} #1\\
  \textsc{Annotation:}\ #2\end{quotation}}

\bibitem[{Abadi et~al.(2015)Abadi, Agarwal, Barham, Brevdo, Chen, Citro
  et~al.}]{TensorFlow15}
[Dataset] Abadi, M., Agarwal, A., Barham, P., Brevdo, E., Chen, Z., Citro, C.,
  et~al. (2015).
\newblock {TensorFlow}: Large-scale machine learning on heterogeneous systems.
\newblock Software available from tensorflow.org
\bibAnnoteFile{TensorFlow15}

\bibitem[{Arridge et~al.(2011)Arridge, André, Khurana, Russell, Cowley, Provan
  et~al.}]{Arridge11}
Arridge, C.~S., André, N., Khurana, K.~K., Russell, C.~T., Cowley, S. W.~H.,
  Provan, G., et~al. (2011).
\newblock Periodic motion of saturn's nightside plasma sheet.
\newblock \emph{Journal of Geophysical Research: Space Physics} 116.
\newblock \doi{10.1029/2011JA016827}
\bibAnnoteFile{Arridge11}

\bibitem[{Azari et~al.(2020)Azari, Biersteker, Dewey, Doran, Forsberg, Harris
  et~al.}]{Azari20}
Azari, A.~R., Biersteker, J.~B., Dewey, R.~M., Doran, G., Forsberg, E.~J.,
  Harris, C. D.~K., et~al. (2020).
\newblock Integrating machine learning for planetary science: Perspectives for
  the next decade.
\newblock \emph{White Paper Submitted to the Decadal Survey on Planetary
  Science and Astrobiology 2023-2032}
\bibAnnoteFile{Azari20}

\bibitem[{Bagenal et~al.(2017)Bagenal, Adriani, Allegrini, Bolton, Bonfond,
  Bunce et~al.}]{Bagenal17}
Bagenal, F., Adriani, A., Allegrini, F., Bolton, S.~J., Bonfond, B., Bunce,
  E.~J., et~al. (2017).
\newblock Magnetospheric science objectives of the juno mission.
\newblock \emph{Space Science Reviews} 213, 219--287.
\newblock \doi{10.1007/s11214-014-0036-8}
\bibAnnoteFile{Bagenal17}

\bibitem[{Bagenal and Delamere(2011)}]{Bagenal11}
Bagenal, F. and Delamere, P.~A. (2011).
\newblock Flow of mass and energy in the magnetospheres of jupiter and saturn.
\newblock \emph{Journal of Geophysical Research: Space Physics} 116.
\newblock \doi{10.1029/2010JA016294}
\bibAnnoteFile{Bagenal11}

\bibitem[{Buda et~al.(2017)Buda, Maki, and Mazurowski}]{Buda17}
Buda, M., Maki, A., and Mazurowski, M.~A. (2017).
\newblock A systematic study of the class imbalance problem in convolutional
  neural networks.
\newblock \emph{CoRR} abs/1710.05381
\bibAnnoteFile{Buda17}

\bibitem[{Buitinck et~al.(2013)Buitinck, Louppe, Blondel, Pedregosa, Mueller,
  Grisel et~al.}]{Buitinck13}
Buitinck, L., Louppe, G., Blondel, M., Pedregosa, F., Mueller, A., Grisel, O.,
  et~al. (2013).
\newblock {API} design for machine learning software: experiences from the
  scikit-learn project.
\newblock In \emph{ECML PKDD Workshop: Languages for Data Mining and Machine
  Learning}. 108--122
\bibAnnoteFile{Buitinck13}

\bibitem[{Bunce et~al.(2005)Bunce, Cowley, Wright, Coates, Dougherty, Krupp
  et~al.}]{Bunce05}
Bunce, E.~J., Cowley, S. W.~H., Wright, D.~M., Coates, A.~J., Dougherty, M.~K.,
  Krupp, N., et~al. (2005).
\newblock In situ observations of a solar wind compression-induced hot plasma
  injection in saturn's tail.
\newblock \emph{Geophysical Research Letters} 32.
\newblock \doi{10.1029/2005GL022888}
\bibAnnoteFile{Bunce05}

\bibitem[{Burkholder et~al.(2017)Burkholder, Delamere, Ma, Thomsen, Wilson, and
  Bagenal}]{Burkholder17}
Burkholder, B., Delamere, P.~A., Ma, X., Thomsen, M.~F., Wilson, R.~J., and
  Bagenal, F. (2017).
\newblock Local time asymmetry of saturn's magnetosheath flows.
\newblock \emph{Geophysical Research Letters} 44, 5877--5883.
\newblock \doi{10.1002/2017GL073031}
\bibAnnoteFile{Burkholder17}

\bibitem[{Cowley et~al.(2015)Cowley, Nichols, and Jackman}]{Cowley15}
Cowley, S. W.~H., Nichols, J.~D., and Jackman, C.~M. (2015).
\newblock Down-tail mass loss by plasmoids in jupiter's and saturn's
  magnetospheres.
\newblock \emph{Journal of Geophysical Research: Space Physics} 120,
  6347--6356.
\newblock \doi{10.1002/2015JA021500}
\bibAnnoteFile{Cowley15}

\bibitem[{{Delamere} et~al.(2015){Delamere}, {Otto}, {Ma}, {Bagenal}, and
  {Wilson}}]{Delamere15}
{Delamere}, P.~A., {Otto}, A., {Ma}, X., {Bagenal}, F., and {Wilson}, R.~J.
  (2015).
\newblock {Magnetic flux circulation in the rotationally driven giant
  magnetospheres}.
\newblock \emph{Journal of Geophysical Research (Space Physics)} 120,
  4229--4245.
\newblock \doi{10.1002/2015JA021036}
\bibAnnoteFile{Delamere15}

\bibitem[{Dougherty et~al.(2004)Dougherty, Kellock, Southwood, Balogh, Smith,
  Tsurutani et~al.}]{Dougherty04}
Dougherty, M., Kellock, S., Southwood, D., Balogh, A., Smith, E., Tsurutani,
  B., et~al. (2004).
\newblock The cassini magnetic field investigation.
\newblock \emph{Space Science Reviews} 114, 331--383.
\newblock \doi{10.1007/s11214-004-1432-2}
\bibAnnoteFile{Dougherty04}

\bibitem[{Dougherty et~al.(2006)Dougherty, Khurana, Neubauer, Russell, Saur,
  Leisner et~al.}]{Dougherty06}
Dougherty, M.~K., Khurana, K.~K., Neubauer, F.~M., Russell, C.~T., Saur, J.,
  Leisner, J.~S., et~al. (2006).
\newblock Identification of a dynamic atmosphere at enceladus with the cassini
  magnetometer.
\newblock \emph{Science} 311, 1406--1409.
\newblock \doi{10.1126/science.1120985}
\bibAnnoteFile{Dougherty06}

\bibitem[{Dungey(1961)}]{Dungey61}
Dungey, J.~W. (1961).
\newblock Interplanetary magnetic field and the auroral zones.
\newblock \emph{Phys. Rev. Lett.} 6, 47--48.
\newblock \doi{10.1103/PhysRevLett.6.47}
\bibAnnoteFile{Dungey61}

\bibitem[{{Dungey}(1965)}]{Dungey65}
{Dungey}, J.~W. (1965).
\newblock {The Length of the Magnetospheric Tail}.
\newblock \emph{Journal of Geophysical Research} 70, 1753--1753.
\newblock \doi{10.1029/JZ070i007p01753}
\bibAnnoteFile{Dungey65}

\bibitem[{Fawaz et~al.(2018)Fawaz, Forestier, Weber, Idoumghar, and
  Muller}]{Fawaz18}
Fawaz, H.~I., Forestier, G., Weber, J., Idoumghar, L., and Muller, P. (2018).
\newblock Data augmentation using synthetic data for time series classification
  with deep residual networks.
\newblock \emph{CoRR} abs/1808.02455
\bibAnnoteFile{Fawaz18}

\bibitem[{Garton(2020)}]{Garton20}
Garton, T.~M. (2020).
\newblock Machine learning identification of reconnection in cassini mag data
  programs \doi{10.5281/zenodo.3978252}
\bibAnnoteFile{Garton20}

\bibitem[{Guo et~al.(2018)Guo, Yao, Wei, Ray, Rae, Arridge et~al.}]{Guo18}
Guo, R.~L., Yao, Z.~H., Wei, Y., Ray, L.~C., Rae, I.~J., Arridge, C.~S., et~al.
  (2018).
\newblock Rotationally driven magnetic reconnection in saturn's dayside.
\newblock \emph{Nature Astronomy} 2, 640--645.
\newblock \doi{10.1038/s41550-018-0461-9}
\bibAnnoteFile{Guo18}

\bibitem[{Guo et~al.(2008)Guo, Yin, Dong, Yang, and Zhou}]{Gua08}
Guo, X., Yin, Y., Dong, C., Yang, G., and Zhou, G. (2008).
\newblock On the class imbalance problem.
\newblock \emph{Fourth International Conference on Natural Computation, ICNC
  '08} Vol. 4.
\newblock \doi{10.1109/ICNC.2008.871}
\bibAnnoteFile{Gua08}

\bibitem[{Heidke(1926)}]{Heidke26}
Heidke, P. (1926).
\newblock Berechnung des erfolges und der güte der windstärkevorhersagen im
  sturmwarnungsdienst.
\newblock \emph{Geografiska Annaler} 8, 301--349.
\newblock \doi{10.1080/20014422.1926.11881138}
\bibAnnoteFile{Heidke26}

\bibitem[{Hill et~al.(2008)Hill, Thomsen, Henderson, Tokar, Coates, McAndrews
  et~al.}]{Hill08}
Hill, T.~W., Thomsen, M.~F., Henderson, M.~G., Tokar, R.~L., Coates, A.~J.,
  McAndrews, H.~J., et~al. (2008).
\newblock Plasmoids in saturn's magnetotail.
\newblock \emph{Journal of Geophysical Research: Space Physics} 113.
\newblock \doi{10.1029/2007JA012626}
\bibAnnoteFile{Hill08}

\bibitem[{{Hones}(1977)}]{Hones77}
{Hones}, J., E.~W. (1977).
\newblock {Substorm processes in the magnetotail: Comments on
  {\textquoteleft}On hot tenuous plasmas, fireballs, and boundary layers in the
  Earth's magnetotail{\textquoteright} by L. A. Frank, K. L. Ackerson, and R.
  P. Lepping}.
\newblock \emph{Journal of Geophysical Research} 82, 5633.
\newblock \doi{10.1029/JA082i035p05633}
\bibAnnoteFile{Hones77}

\bibitem[{Huang(2003)}]{Huang03}
Huang, G.-B. (2003).
\newblock Huang, g.: Learning capability and storage capacity of
  two-hidden-layer feedforward networks. ieee trans. on neural networks 14(2),
  274-281.
\newblock \emph{IEEE transactions on neural networks / a publication of the
  IEEE Neural Networks Council} 14, 274--81.
\newblock \doi{10.1109/TNN.2003.809401}
\bibAnnoteFile{Huang03}

\bibitem[{Jackman et~al.(2013)Jackman, Achilleos, Cowley, Bunce, Radioti,
  Grodent et~al.}]{Jackman13}
Jackman, C.~M., Achilleos, N., Cowley, S.~W., Bunce, E.~J., Radioti, A.,
  Grodent, D., et~al. (2013).
\newblock Auroral counterpart of magnetic field dipolarizations in saturn's
  tail.
\newblock \emph{Planetary and Space Science} 82-83, 34 -- 42.
\newblock \doi{https://doi.org/10.1016/j.pss.2013.03.010}
\bibAnnoteFile{Jackman13}

\bibitem[{Jackman et~al.(2009)Jackman, Arridge, McAndrews, Henderson, and
  Wilson}]{Jackman09}
Jackman, C.~M., Arridge, C.~S., McAndrews, H.~J., Henderson, M.~G., and Wilson,
  R.~J. (2009).
\newblock Northward field excursions in saturn's magnetotail and their
  relationship to magnetospheric periodicities.
\newblock \emph{Geophysical Research Letters} 36.
\newblock \doi{10.1029/2009GL039149}
\bibAnnoteFile{Jackman09}

\bibitem[{Jackman et~al.(2007)Jackman, Russell, Southwood, Arridge, Achilleos,
  and Dougherty}]{Jackman07}
Jackman, C.~M., Russell, C.~T., Southwood, D.~J., Arridge, C.~S., Achilleos,
  N., and Dougherty, M.~K. (2007).
\newblock Strong rapid dipolarizations in saturn's magnetotail: In situ
  evidence of reconnection.
\newblock \emph{Geophysical Research Letters} 34.
\newblock \doi{10.1029/2007GL029764}
\bibAnnoteFile{Jackman07}

\bibitem[{Jackman et~al.(2011)Jackman, Slavin, and Cowley}]{Jackman11}
Jackman, C.~M., Slavin, J.~A., and Cowley, S. W.~H. (2011).
\newblock Cassini observations of plasmoid structure and dynamics: Implications
  for the role of magnetic reconnection in magnetospheric circulation at
  saturn.
\newblock \emph{Journal of Geophysical Research: Space Physics} 116.
\newblock \doi{10.1029/2011JA016682}
\bibAnnoteFile{Jackman11}

\bibitem[{Jackman et~al.(2014)Jackman, Slavin, Kivelson, Southwood, Achilleos,
  Thomsen et~al.}]{Jackman14}
Jackman, C.~M., Slavin, J.~A., Kivelson, M.~G., Southwood, D.~J., Achilleos,
  N., Thomsen, M.~F., et~al. (2014).
\newblock Saturn's dynamic magnetotail: A comprehensive magnetic field and
  plasma survey of plasmoids and traveling compression regions and their role
  in global magnetospheric dynamics.
\newblock \emph{Journal of Geophysical Research: Space Physics} 119,
  5465--5494.
\newblock \doi{10.1002/2013JA019388}
\bibAnnoteFile{Jackman14}

\bibitem[{Jackman et~al.(2015)Jackman, Thomsen, Mitchell, Sergis, Arridge,
  Felici et~al.}]{Jackman15}
Jackman, C.~M., Thomsen, M.~F., Mitchell, D.~G., Sergis, N., Arridge, C.~S.,
  Felici, M., et~al. (2015).
\newblock Field dipolarization in saturn's magnetotail with planetward ion
  flows and energetic particle flow bursts: Evidence of quasi-steady
  reconnection.
\newblock \emph{Journal of Geophysical Research: Space Physics} 120,
  3603--3617.
\newblock \doi{10.1002/2015JA020995}
\bibAnnoteFile{Jackman15}

\bibitem[{Johnson and Khoshgoftaar(2019)}]{Johnson19}
Johnson, J.~M. and Khoshgoftaar, T.~M. (2019).
\newblock Survey on deep learning with class imbalance.
\newblock \emph{Journal of Big Data} 6, 27.
\newblock \doi{10.1186/s40537-019-0192-5}
\bibAnnoteFile{Johnson19}

\bibitem[{Kane et~al.(2020)Kane, Mitchell, Carbary, Dialynas, Hill, and
  Krimigis}]{Kane20}
Kane, M., Mitchell, D.~G., Carbary, J.~F., Dialynas, K., Hill, M.~E., and
  Krimigis, S.~M. (2020).
\newblock Convection in the magnetosphere of saturn during the cassini mission
  derived from mimi inca and chems measurements.
\newblock \emph{Journal of Geophysical Research: Space Physics} 125,
  e2019JA027534.
\newblock \doi{10.1029/2019JA027534}.
\newblock E2019JA027534 2019JA027534
\bibAnnoteFile{Kane20}

\bibitem[{Khurana et~al.(2009)Khurana, Mitchell, Arridge, Dougherty, Russell,
  Paranicas et~al.}]{Khurana09}
Khurana, K.~K., Mitchell, D.~G., Arridge, C.~S., Dougherty, M.~K., Russell,
  C.~T., Paranicas, C., et~al. (2009).
\newblock Sources of rotational signals in saturn's magnetosphere.
\newblock \emph{Journal of Geophysical Research: Space Physics} 114.
\newblock \doi{10.1029/2008JA013312}
\bibAnnoteFile{Khurana09}

\bibitem[{Ma et~al.(2017)Ma, Delamere, Otto, and Burkholder}]{Ma17}
Ma, X., Delamere, P., Otto, A., and Burkholder, B. (2017).
\newblock Plasma transport driven by the three-dimensional kelvin-helmholtz
  instability.
\newblock \emph{Journal of Geophysical Research: Space Physics} 122,
  10,382--10,395.
\newblock \doi{10.1002/2017JA024394}
\bibAnnoteFile{Ma17}

\bibitem[{Martin and Arridge(2017)}]{Martin17}
Martin, C.~J. and Arridge, C.~S. (2017).
\newblock Cassini observations of aperiodic waves on saturn's magnetodisc.
\newblock \emph{Journal of Geophysical Research: Space Physics} 122,
  8063--8077.
\newblock \doi{10.1002/2017JA024293}
\bibAnnoteFile{Martin17}

\bibitem[{McAndrews et~al.(2009)McAndrews, Thomsen, Arridge, Jackman, Wilson,
  Henderson et~al.}]{Mcandrews07}
McAndrews, H., Thomsen, M., Arridge, C., Jackman, C., Wilson, R., Henderson,
  M., et~al. (2009).
\newblock Plasma in saturn's nightside magnetosphere and the implications for
  global circulation.
\newblock \emph{Planetary and Space Science} 57, 1714 -- 1722.
\newblock \doi{https://doi.org/10.1016/j.pss.2009.03.003}
\bibAnnoteFile{Mcandrews07}

\bibitem[{Mikołajczyk and Grochowski(2018)}]{Mikolajczyk18}
Mikołajczyk, A. and Grochowski, M. (2018).
\newblock Data augmentation for improving deep learning in image classification
  problem.
\newblock 117--122.
\newblock \doi{10.1109/IIPHDW.2018.8388338}
\bibAnnoteFile{Mikolajczyk18}

\bibitem[{Milan et~al.(2007)Milan, Provan, and Hubert}]{Milan07}
Milan, S.~E., Provan, G., and Hubert, B. (2007).
\newblock Magnetic flux transport in the dungey cycle: A survey of dayside and
  nightside reconnection rates.
\newblock \emph{Journal of Geophysical Research: Space Physics} 112.
\newblock \doi{10.1029/2006JA011642}
\bibAnnoteFile{Milan07}

\bibitem[{Nakagawa and Nishida(1989)}]{Nakagawa89}
Nakagawa, T. and Nishida, A. (1989).
\newblock Southward magnetic field in the neutral sheet produced by wavy
  motions propagating in the dawn-dusk direction.
\newblock \emph{Geophysical Research Letters} 16, 1265--1268.
\newblock \doi{10.1029/GL016i011p01265}
\bibAnnoteFile{Nakagawa89}

\bibitem[{Neupane et~al.(2019)Neupane, Delamere, Wilson, and Ma}]{Neupane19}
Neupane, B.~R., Delamere, P.~A., Wilson, R.~J., and Ma, X. (2019).
\newblock Quantifying mass and magnetic flux transport in saturn's
  magnetosphere.
\newblock \emph{Journal of Geophysical Research: Space Physics} 124,
  1916--1926.
\newblock \doi{10.1029/2018JA026022}
\bibAnnoteFile{Neupane19}

\bibitem[{{Ruhunusiri}(2018)}]{Ruhunusiri18b}
{Ruhunusiri}, S. (2018).
\newblock Identification of plasma waves at saturn using convolutional neural
  networks.
\newblock \emph{IEEE Transactions on Plasma Science} 46, 3090--3099
\bibAnnoteFile{Ruhunusiri18b}

\bibitem[{Ruhunusiri et~al.(2018)Ruhunusiri, Halekas, Espley, Eparvier, Brain,
  Mazelle et~al.}]{Ruhunusiri18}
Ruhunusiri, S., Halekas, J.~S., Espley, J.~R., Eparvier, F., Brain, D.,
  Mazelle, C., et~al. (2018).
\newblock An artificial neural network for inferring solar wind proxies at
  mars.
\newblock \emph{Geophysical Research Letters} 45, 10,855--10,865.
\newblock \doi{10.1029/2018GL079282}
\bibAnnoteFile{Ruhunusiri18}

\bibitem[{Russell et~al.(2008)Russell, Jackman, Wei, Bertucci, and
  Dougherty}]{Russell08}
Russell, C.~T., Jackman, C.~M., Wei, H.~Y., Bertucci, C., and Dougherty, M.~K.
  (2008).
\newblock Titan’s influence on saturnian substorm occurrence.
\newblock \emph{Geophysical Research Letters} 35.
\newblock \doi{10.1029/2008GL034080}
\bibAnnoteFile{Russell08}

\bibitem[{{Simon} et~al.(2010){Simon}, {Wennmacher}, {Neubauer}, {Bertucci},
  {Kriegel}, {Saur} et~al.}]{Sven10}
{Simon}, S., {Wennmacher}, A., {Neubauer}, F.~M., {Bertucci}, C.~L., {Kriegel},
  H., {Saur}, J., et~al. (2010).
\newblock {Titan's highly dynamic magnetic environment: A systematic survey of
  Cassini magnetometer observations from flybys TA-T62}.
\newblock \emph{Planetary and Space Science} 58, 1230--1251.
\newblock \doi{10.1016/j.pss.2010.04.021}
\bibAnnoteFile{Sven10}

\bibitem[{Slavin et~al.(1984)Slavin, Smith, Tsurutani, Sibeck, Singer, Baker
  et~al.}]{Slavin84}
Slavin, J.~A., Smith, E.~J., Tsurutani, B.~T., Sibeck, D.~G., Singer, H.~J.,
  Baker, D.~N., et~al. (1984).
\newblock Substorm associated traveling compression regions in the distant
  tail: Isee-3 geotail observations.
\newblock \emph{Geophysical Research Letters} 11, 657--660.
\newblock \doi{10.1029/GL011i007p00657}
\bibAnnoteFile{Slavin84}

\bibitem[{Smith et~al.(2016)Smith, Jackman, and Thomsen}]{Smith15}
Smith, A.~W., Jackman, C.~M., and Thomsen, M.~F. (2016).
\newblock Magnetic reconnection in saturn's magnetotail: A comprehensive
  magnetic field survey.
\newblock \emph{Journal of Geophysical Research: Space Physics} 121,
  2984--3005.
\newblock \doi{10.1002/2015JA022005}
\bibAnnoteFile{Smith15}

\bibitem[{Smith et~al.(2018{\natexlab{a}})Smith, Jackman, Thomsen, Lamy, and
  Sergis}]{Smith18a}
Smith, A.~W., Jackman, C.~M., Thomsen, M.~F., Lamy, L., and Sergis, N.
  (2018{\natexlab{a}}).
\newblock Multi-instrument investigation of the location of saturn's
  magnetotail x-line.
\newblock \emph{Journal of Geophysical Research: Space Physics} 123,
  5494--5505.
\newblock \doi{10.1029/2018JA025532}
\bibAnnoteFile{Smith18a}

\bibitem[{Smith et~al.(2018{\natexlab{b}})Smith, Jackman, Thomsen, Sergis,
  Mitchell, and Roussos}]{Smith18b}
Smith, A.~W., Jackman, C.~M., Thomsen, M.~F., Sergis, N., Mitchell, D.~G., and
  Roussos, E. (2018{\natexlab{b}}).
\newblock Dipolarization fronts with associated energized electrons in saturn's
  magnetotail.
\newblock \emph{Journal of Geophysical Research: Space Physics} 123,
  2714--2735.
\newblock \doi{10.1002/2017JA024904}
\bibAnnoteFile{Smith18b}

\bibitem[{Solomon et~al.(2001)Solomon, McNutt, Gold, Acuña, Baker, Boynton
  et~al.}]{Solomon01}
Solomon, S., McNutt, R., Gold, R., Acuña, M., Baker, D., Boynton, W., et~al.
  (2001).
\newblock The messenger mission to mercury: Scientific objectives and
  implementation.
\newblock \emph{Planetary and Space Science} 49, 1445--1465.
\newblock \doi{10.1016/S0032-0633(01)00085-X}
\bibAnnoteFile{Solomon01}

\bibitem[{Thomsen et~al.(2013)Thomsen, Wilson, Tokar, Reisenfeld, and
  Jackman}]{Thomsen13}
Thomsen, M.~F., Wilson, R.~J., Tokar, R.~L., Reisenfeld, D.~B., and Jackman,
  C.~M. (2013).
\newblock Cassini/caps observations of duskside tail dynamics at saturn.
\newblock \emph{Journal of Geophysical Research: Space Physics} 118,
  5767--5781.
\newblock \doi{10.1002/jgra.50552}
\bibAnnoteFile{Thomsen13}

\bibitem[{Vasyliunas(1983)}]{Vasyliunas83}
Vasyliunas, V.~M. (1983).
\newblock \emph{Plasma distribution and flow} (Cambridge University Press).
\newblock Cambridge Planetary Science Old. 395–453.
\newblock \doi{10.1017/CBO9780511564574.013}
\bibAnnoteFile{Vasyliunas83}

\bibitem[{Waldmann and Griffith(2019)}]{Waldmann19}
Waldmann, I.~P. and Griffith, C.~A. (2019).
\newblock Mapping saturn using deep learning.
\newblock \emph{Nature Astronomy} 3, 620--625.
\newblock \doi{10.1038/s41550-019-0753-8}
\bibAnnoteFile{Waldmann19}

\bibitem[{Yao et~al.(2017)Yao, Grodent, Ray, Rae, Coates, Pu et~al.}]{Yao17}
Yao, Z.~H., Grodent, D., Ray, L.~C., Rae, I.~J., Coates, A.~J., Pu, Z.~Y.,
  et~al. (2017).
\newblock Two fundamentally different drivers of dipolarizations at saturn.
\newblock \emph{Journal of Geophysical Research: Space Physics} 122,
  4348--4356.
\newblock \doi{10.1002/2017JA024060}
\bibAnnoteFile{Yao17}

\bibitem[{Young et~al.(2004)Young, Berthelier, Blanc, Burch, Coates, Goldstein
  et~al.}]{Young04}
Young, D.~T., Berthelier, J.~J., Blanc, M., Burch, J.~L., Coates, A.~J.,
  Goldstein, R., et~al. (2004).
\newblock Cassini plasma spectrometer investigation.
\newblock \emph{Space Science Reviews} 114, 1--112.
\newblock \doi{10.1007/s11214-004-1406-4}
\bibAnnoteFile{Young04}

\bibitem[{Young(1998)}]{Young98}
Young, R.~E. (1998).
\newblock The galileo probe mission to jupiter: Science overview.
\newblock \emph{Journal of Geophysical Research: Planets} 103, 22775--22790.
\newblock \doi{10.1029/98JE01051}
\bibAnnoteFile{Young98}

\end{thebibliography}


\section*{Figure captions}

\end{document}